# Progress in artificial intelligence applications based on the combination of self-driven sensors and deep learning

Weixiang Wan [1,*]

Electronics & Communication Engineering

University of Electronic Science and Technology of China

Chengdu,China

danielwanwx@gmail.com

Qiang Zeng [3]

Software Engineering

Zhejiang University

Hangzhou, Zhejiang, China

zengqiang@zju.edu.cn

Bo Liu [6]

Software Engineering

Zhejiang University

Hangzhou, Zhejiang, China

21851111@zju.edu.cn

Wenjian Sun [2]

Electronic and Information Engineering

Yantai University

Tokoy,Japan

swjhuman@gmail.com

Linying Pan [4]

Information Studies

Trine university

Phoenix, Arizona, USA

panlinying2023@gmail.com

Jingyu Xu [5]

Computer Information Technology

Northern Arizona University

Flagstaff,Arizona,USA

jyxu01@outlook.com

*Abstract*—In the era of Internet of Things, how to develop a smart sensor system with sustainable power supply, easy deployment and flexible use has become a difficult problem to be solved. The traditional power supply has problems such as frequent replacement or charging when in use, which limits the development of wearable devices. The contact-to-separate friction nanogenerator (TENG) was prepared by using polychotomy thy lene (PTFE) and aluminum (AI) foils. Human motion energy was collected by human body arrangement, and human motion posture was monitored according to the changes of output electrical signals. In 2012, Academician Wang Zhong lin and his team invented the triboelectric nanogenerator (TENG), which uses Maxwell displacement current as a driving force to directly convert mechanical stimuli into electrical signals, so it can be used as a self-driven sensor. Teng-based sensors have the advantages of simple structure and high instantaneous power density, which provides an important means for building intelligent sensor systems. At the same time, machine learning, as a technology with low cost, short development cycle, strong data processing ability and prediction ability, has a significant effect on the processing of a large number of electrical signals generated by TENG, and the combination with TENG sensors will promote the rapid development of intelligent sensor networks in the future. Therefore, this paper is based on the intelligent sound monitoring and recognition system of TENG, which has good sound recognition capability, and aims to evaluate the feasibility of the sound perception module architecture in ubiquitous sensor networks. Transportation, security, water conservancy, construction and other fields need urban sound management. Given the variety of noise sensors available and the wide range of opportunities to connect these sensors through mobile broadband Internet access, many researchers are eager to apply sound sensor networks to urban sound management. Existing sensor networks often consist of expensive information sensing devices whose cost and maintenance limit their large-scale deployment, thereby narrowing their functional measurement range.

I. INTRODUCTION (HEADING 1)

In recent years, with the rapid development of the Internet of Things, human needs for information storage, processing and use methods have become increasingly prominent. At the heart of the Internet of Things is a wireless sensor network made up of billions of sensors that are widely distributed and can help people monitor their surroundings and human health. With the promotion and application of the Internet of Things, the number of wireless sensor nodes has increased rapidly, and the problems such as the power supply and use flexibility of sensors have become more and more prominent, which directly leads to the problems of short service life, difficult node deployment, and poor maintainability of traditional sensor networks[1]. Traditional sensors are powered by the grid and batteries, and the energy supply will face huge challenges, and the demand for sensors in the future Internet of Things is still increasing. Therefore, a self-powered sensor that does not require a power supply is the most promising and sustainable solution to this problem. TENG has a wide variety of sensor types and a wide range of applications, including self-actuated pressure sensors. Self-actuated sound sensor, self-actuated acceleration sensor, self-actuated temperature sensor, self-actuated humidity sensor, self-actuated gas sensor, etc. These self-actuated sensors are designed to partially replace traditional sensors to save energy and power consumption. Therefore, TENG, as a new sensor technology, has a broader application prospect.

Secondly, due to the continuous optimization and development of the environment of machine learning and artificial intelligence in recent years, especially deep learning has gradually become a research hotspot and mainstream development direction in the field of Internet of Things and sensors, the combination of big data and machine learning technology and TENG sensor data processing provides a new idea for researchers. Machine learning was proposed by Arthur Samuel in 1959 and has been widely used in computer vision, general games, economics, data mining, and bioinformatics. Artificial intelligence is an interdisciplinary field of study. In recent years, with the improvement of artificial intelligence and machine learning theory and the expansion of tool chain, more and more experts have made new achievements in their research field with the help of this powerful tool[2]. Because machine learning resources and tools are abundant and easy to obtain, it has more efficient data processing capabilities, which can make the sensor intelligent and save test time and cost. Compared with traditional sensors, data generated by TENG sensors is more suitable for processing and analysis by machine learning technology. Therefore, machine learning has gradually appeared in TENG's research field.

## II. RELATED THEORETCAL KNOWLED

### A. TENG principle

The TENG is an energy harvesting and conversion device that converts small scale mechanical energy in the environment into electrical energy at the nanoscale. TENG has significant advantages: no pollution, can collect and convert energy in the working environment, small size, simple structure, easy to manufacture, high pressure characteristics, etc. The basic principles of TENG are triboelectrification and electrostatic induction. Although the phenomenon of triboelectrification has existed in ancient times, the mechanism of electrification is still unclear. One explanation is widely accepted: When two materials come into contact with each other, a small amount of chemical bond forms on the surface, which transfers charges and balances the potential difference. When the two materials separate, some bonding atoms give away electrons, and some bonding atoms retain electrons due to differences in their ability to gain and lose electrons, and the surface of the two friction layers generates friction charges[3-5]. This frictional charge can create a potential difference in an external circuit.

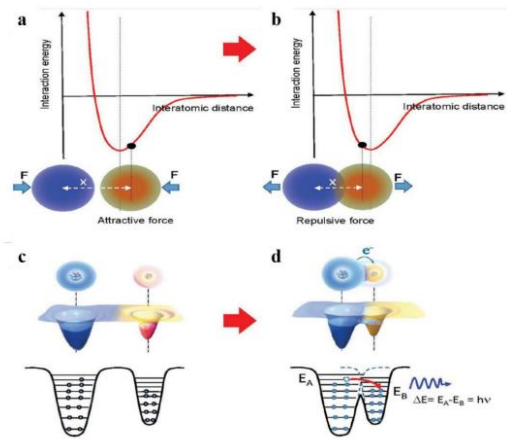

Figure 1. Charge transfer model of TENG

As a sustainable power supply device, TENG can collect and convert energy in the working environment, reduce the use of batteries, and provide a new idea for the management of potential environmental risks of batteries. The TENG also has the potential to capture and transform mechanical energy found in nature on a large scale, promising a new avenue for blue energy supply[6]. In addition, the energy obtained from the working environment and converted into electrical signals makes the TENG have unparalleled application advantages in some special environments. The self-powered sensor developed based on the principles of the TENG can be applied in a variety of scenarios, such as human-computer interaction, environmental monitoring, and healthcare. TENG has shown broad application prospects in clean energy, self-powered sensing and so on

### B. TENG working mode

The basic principles of TENG are triboelectrification and electrostatic induction. Although the phenomenon of triboelectrification has existed in ancient times, the mechanism of electrification is still unclear. One explanation is widely accepted: When two materials come into contact with each other, a small amount of chemical bond forms on the surface, which transfers charges and balances the potential difference. When the two materials separate, some bonding atoms give away electrons, and some bonding atoms retain electrons due to differences in their ability to gain and lose electrons, and the surface of the two friction layers generates friction charges. This friction charge can create a potential difference in an external circuit. The four common operating modes of TENG are 1, vertical contact separation type 2, lateral sliding type 3, single electrode type 4, and independent friction layer type.

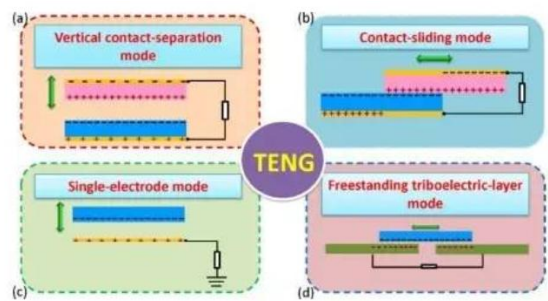

Figure 2. Four basic working modes of triboelectric nanogenerator

The vertical contact separated TENG(a) is mainly composed of two parts: triboelectric generation layer and electrostatic induction layer. The triboelectric generation layer is a dielectric with large electronegativity difference between the two layers, and the electrostatic induction layer is the electrode on the back of the dielectric. The difference between the transverse sliding TENG(b) and the vertical contact separated TENG lies in the different motion modes of the friction layer. The transverse sliding TENG generates a potential difference through the relative sliding of the friction layer and achieves charge balance in the external circuit, generating an electric current in the process. It should be emphasized that sliding can be planar or rotational sliding; Single-electrode TENG(c) has only the bottom electrode and is grounded. The local electric field distribution can be changed by adjusting the distance between the charged friction layer at the top and the bottom electrode, which makes the bottom electrode and the ground exchange electrons, thus generating current. Under the friction layer of the independent friction layer TENG(d), there are two electrodes of the same size as the friction layer. When the charged friction layer moves over the two electrodes, the balanced charge distribution between the two electrodes will be disrupted, thus generating current in the external circuit. Usually, moving objects will be charged due to air friction and other reasons, and this static electricity will exist for a long period of time without contact with other objects. In practical applications, since TENG is mostly used in motion scenes, it may not be convenient to connect electrodes, so single-electrode TENG is more suitable[7-9].

C. TENG evaluation method

The four operating modes of the TENG adapt to different mechanical triggering conditions, which makes it difficult to evaluate the performance of the TENG in different modes with a uniform quantitative method. Based on the cumulative voltage-transfer charge curve, Zi et al. proposed a method to quantitatively evaluate the properties of TENG from both structural and material perspectives. From the structural point of view, the results of theoretical calculation and finite element analysis show that the paired electrodes, the trigger mode of contact separation and the design of independent layer structure are helpful to enhance the performance of TENG. The structural factor of quality (FOMs) is used as the criterion to evaluate the structure, which is derived from the maximum FOMs extracted in the finite element simulation.

$$FOM_S = \frac{2\varepsilon_0}{\sigma^2} \frac{E_m}{Ax_{max}} \quad (1)$$

In the formula, $E_m$ is the maximum possible output energy per cycle, μJ; $x_{max}$ is the maximum displacement of the friction pair, m; σ is the friction surface charge density, μC/m2; ε0 is the dielectric constant of vacuum (ε0=8.85×10-12 F/m); A is the contact area, m2

III. MACHINE LEARNING COMBINED WITH TENG

At present, the most important uses of the TENG are two: as an energy source for sensors and as a self-actuated sensor. The current TENG-based sensors include sound sensors, pressure sensors, moving object self-actuated sensors, vibration sensors, self-actuated environmental sensors, and biomedical sensors. The data that can be collected generally includes audio information, image information, graphic information, amplitude information, spatial position information, temperature information and frequency information, etc[10]. The data sources are complex and the forms are diverse. With the rapid development of machine learning technology, more and more people are using machine learning technology to process massive and complex sensor data. Machine learning, a branch of artificial intelligence, is a technique of building automatic learning models and using large amounts of data to continuously optimize the models, and then making sound predictions guided by algorithms. Machine learning workflows can be simply described as data preprocessing, feature extraction, modeling, validation, and output results. Generally speaking, the use of machine learning technology for training and recognition generally requires data preprocessing, feature extraction, selection of appropriate machine learning methods and establishment of models, and result verification. The general process of using the machine learning method to process data generated by the TENG sensor is that the sensor collects the data and transmits it to the data center for storage and management. Then, the user sends the selected sensor data to the machine learning model through the control center and classifies the data in the data preprocessing stage of the model. It is generally divided into multivariate time series and multivariate spatial data[11]. Then feature extraction and normalization of the classified data are carried out, and machine learning methods are used for analysis and prediction. Finally, the results are verified and evaluated by the machine learning model, and the user can also store and display the results through the control center

A. Data pre-processing

Data preprocessing means that before the data is trained and learned, the format of the data needs to be unified according to the requirements of the machine learning algorithm. The methods of data preprocessing mainly include data cleaning, normalization, feature discovery, feature selection and unbalanced data management. Data clarity is the process of detecting and correcting inaccurate data. Feature engineering is the process of using domain knowledge of data to create machine learning algorithms that work. Feature selection, also known as variable selection/attribute selection, is the process of selecting a subset of relevant features for use in model construction. Good data preprocessing can improve the quality of data, so that the model can learn the data better, and then determine the prediction and generalization ability of the model.

B. Feature extraction

After data collection and normalization, appropriate features should be extracted to predict the target, which is called feature selection level data preprocessing, and feature extraction can be realized by calculating the defined features or using machine learning technology. For low-dimensional data, feature extraction can be omitted. However, with the trend of developing high-density and fast sampling rate triboelectric sensors, many high-dimensional triboelectric signals have emerged, and the need for feature extraction has emerged.

Feature selection and corresponding feature extraction methods are very important for the accuracy of the final result. Reasonable selection can not only retain most of the information of the signal, but also minimize data redundancy, and then input the processed data into the machine learning algorithm to achieve the intended application goal of the sensor. The choice of features and the choice of machine learning algorithms both play an important role in the final result. On the one hand, with proper feature extraction strategies, even simple machine learning algorithms can provide good results. On the other hand, a sufficiently powerful machine learning algorithm can perform satisfactorily even without a feature extraction step. In practice, the combination of these two aspects of careful selection provides researchers with powerful data processing tools.

*C. Machine learning integration*

With the right and enough data and features, you can build a model that analyzes the data according to the requirements. The modeling steps include choosing the right algorithm, training from the trained data, and making accurate predictions. Machine learning tasks can be roughly divided into four categories: classification, regression, clustering, and dimensionality reduction, and their methods can be divided into supervised learning, unsupervised learning, semi-supervised learning, and reinforcement learning. Supervised learning is also called "learning with a teacher" i.e. the corresponding output of the training data has been labeled, where classification and regression are the tasks of supervised learning. In contrast, the corresponding output of the training data in unsupervised learning is unlabeled. For semi-supervised learning, part of the training data is labeled and the rest is not labeled: the amount of unlabeled data usually far exceeds the amount of labeled data[12].

TABLE I. DIFFERENT LEARNING MODES FOR TENG

| | metholgy | |
|---|---|---|
| | Learning method | Learning algorithm |
| TENG | Supervised learning, | SVM |
| | unsupervised learning, | ANN |
| | semi-supervised learning, | CNN |
| | machine learning, | LSTM |
| | reinforcement learning | DBN |

In addition to supervised and unsupervised learning, reinforcement learning is another typical machine learning algorithm. Reinforcement learning emphasizes solving decision problems and is commonly used in robotic operations to calculate the next action. Specific machine learning methods include support vector machine (SVM), artificial neural network (ANN), and convolutional neural network (convolutional neural network) network. CNN), long short-term memory (LSTM), deep beliefnet work (DBN), etc. There are a large number of machine learning algorithms, each with its own unique advantages and scope of application. According to different applications and different data sets, more suitable methods can be selected according to the signal characteristics of triboelectric sensors or the advantages of different methods can be integrated to build a special learning model.

IV. DIFFERENT LEARNING MODES FOR TENG

Drive sensing is a mechanical wave generated by the vibration of an object, propagating through air, water and solid media. Sound plays an important role in our daily life, wherever we go we will be surrounded by sound. Sound can convey information, such as language and music, as well as other features of the environment, such as car noise and bird song. In addition, sound can provide a wealth of information, which has a unique advantage in cases where sight, touch, and smell do not apply. Secondly, sound recognition is an important attribute of sound perception module in ubiquitous sensor network, which is the basis of many high-precision perception and behavior activities. Sound can convey information, such as language and music, as well as other features of the environment, such as car noise and bird song. Therefore, sound monitoring technology utilizing TENG sensors can play an important role in socio-economic life.

*A. Sound sensor*

The first is the sound optimization application of the TENG technology under the self-driven sensor. When the sound in the environment is transmitted to the TENG in the form of sound waves, the PTFE film will periodically vibrate under the action of the pressure difference between the two sides. At the same time, the voltage between the two electrodes changes with the vibration of the PTFE film, achieving the response and recording of sound.

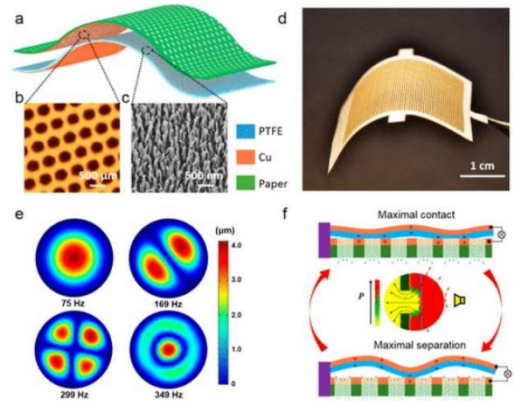

Figure 3. Working mode and structural design of TENG based on ultra-thin paper

Compared with the traditional sound driven friction nanogenerator sensor, this ultra-thin flexible sound sensor has the advantages of high sensitivity, high energy conversion efficiency and good stability.

*B. Intelligent robot application*

To further promote the development of friction nano dynamo-based sound sensors, Guo et al. designed a single-channel, circular, self-powered triboelectric hearing sensor (TAS) to construct an architecture for external hearing AIDS and an electronic hearing system for intelligent robotic applications[13]. Based on the newly developed cryoground nanogenerator (TENG) technology, the TAS shows extremely high sensitivity (110mV/dB). A wide broadband response of 100 to 5000 Hz is achieved, covering almost all frequency ranges of the human voice.

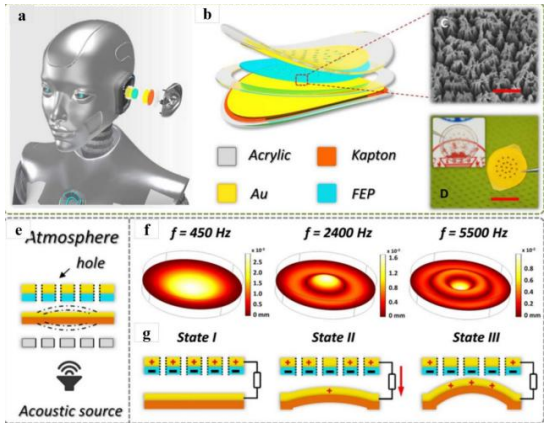

Figure 4. The structure and mechanism of TAS

It proved to be a highly efficient, high-fidelity, low-cost hearing platform for social robot interaction. By adjusting the inner boundary structure of the ring/sector, the custom frequency response of the sensor itself was realized, and the application as a hearing aid device to help the rehabilitation of the hearing-impaired was demonstrated, thus enhancing the interaction between humans and robots audio logically.

*C. Simulated biological hearing*

The self-powered artificial hearing pathway consists of a femtosecond laser-treated TENG and a field-effect synaptic transistor (FEST), which act as acoustic receptors and acoustic synapses, respectively. It is worth mentioning that the femtosecond laser-induced auditory TENG not only exhibited a wider frequency response and high sensitivity (129 mV/dB), but also excelled in empathic social behavior. In addition, various typical synaptic functions can be well simulated through the TENG-driven FEST. At the same time, the study also demonstrates an adaptive artificial neuromorphic circuit with noise tunable behavior, which can greatly improve the efficiency and accuracy of instruction recognition, providing better performance for more human-computer interaction systems.

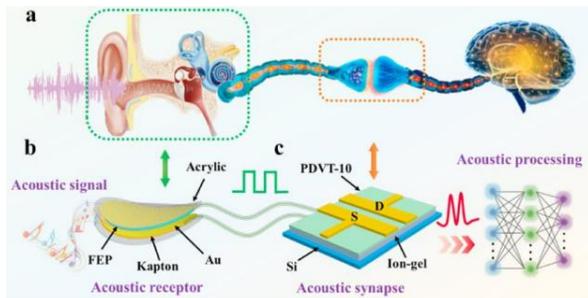

Figure 5. Auditory simulation diagram

This work provides an efficient way to mimic biological hearing functions, achieve instruction recognition in the noise intensity range, and will reduce power consumption in the field of neuromorphic systems in the future.

However, although these efforts have promoted the development of TENG-based sensors in the field of voice recognition, the degree of intelligence of the sensors used still needs to be improved, and the lack of efficient analysis and calculation means is generally difficult to cope with the accurate calculation and processing of large-scale data.

## V. CONCLUSIONS

With the advent of the Internet of Things (IoT) era and the increasing importance of information to the development of human society, there is an increasing demand for large-scale and inter-city iot. How to develop the ubiquitous sensor network (USN) with large-scale regional data collection and intelligent analysis ability is of great significance to People's Daily life needs. In recent years, friction nanogenerators (TENG) have attracted a lot of attention due to their cost-effective, easy to manufacture, easy to maintain and flexible layout, and their ability to convert mechanical motion into electrical signals, making them useful as self-powered sensors. As a technology with short development cycle, powerful data processing ability and strong prediction ability for diverse data sets, deep learning can well solve the problems of large volume, poor continuity and low efficiency of TENG data[15]. Therefore, the use of TENG-based sensors combined with deep learning to develop a cheap and flexible smart sensor system with sustainable power supply is the best solution to achieve large-scale deployment of USN and inter-city iot. Voice recognition is an important attribute of the USN voice perception module and is the basis for many high-precision perception and behavioral activities. Sound signals provide a wealth of information and have unique advantages in situations where sight, touch, and smell do not apply. Therefore, the use of smart sound monitoring technology based on TENG sensors can play an important role in socio-economic life[16].

Nowadays, the appearance of aging population makes the traditional clinical diagnosis method can not meet the diversified medical needs of people. In addition to pathological treatment, people also need to monitor and know their health status at all times. An important aspect of healthcare surveillance is the ability to detect when there are abnormalities that could point to disease. The emergence of the Internet of Things (iot), a new model for collecting and analyzing data anytime, anywhere, is helping to shift healthcare from treatment to prevention. Respiratory monitoring and heart rate monitoring are crucial to human health assessment and play an important role in the field of healthcare[17-18]. The wearable TENG for respiration monitoring and heart rate monitoring is a cheaper and more convenient method for long-term monitoring of human physiological parameter information, which is capable of combining various elements necessary for modern portable healthcare systems, such as Internet of Things, integrated chips, artificial intelligence, disease monitoring and point of care. However, many aspects of these elements still need to be perfected before they can be used commercially. To further improve the user experience, the use of AI algorithms or machine learning is a viable solution, similar to the Internet of Things, which has recently gained popularity.

ACKNOWLEDGMENT

The preferred spelling of the word "acknowledgment" in

## REFERENCES

[1] McCarthy l, Feigenbaum E A. In memoriam: Arthur samuel: Pioneer in machine learning[J],AI Magazine, 1990, 11(3): 10-10

[2] Zhou L, Liu D, Wang J, et al. Triboelectric nanogenerators: fundamental physics and potentialapplicationsJ. Friction, 2020, 8: 481-506.

[3] Lin, Q., Che, C., Hu, H., Zhao, X., & Li, S. (2023). A Comprehensive Study on Early Alzheimer's Disease Detection through Advanced Machine Learning Techniques on MRI Data. Academic Journal of Science and Technology, 8(1), 281–285.DOI: 10.1111/jgs.18617

[4] Luo J, Wang Z L. Recent progress of triboelectric nanogenerators: From fundamental theoryto practical applicationsJ]. EcoMat, 2020, 2(4): e12059.8

[5] Nie J, Ren Z, Xu L, et al. Probing contact-electrification-induced electron and ion transfers ata liquid solid interface J . Advanced Materials, 2020, 32(2): 1905696.

[6] Che, C., Hu, H., Zhao, X., Li, S., & Lin, Q. (2023). Advancing Cancer Document Classification with R andom Forest. Academic Journal of Science and Technology, 8(1), 278–280. https://doi.org/10.54097/ajst.v8i1.14333

[7] Tianbo, Song, Hu Weijun, Cai Jiangfeng, Liu Weijia, Yuan Quan, and He Kun. "Bio-inspired Swarm Intelligence: a Flocking Project With Group Object Recognition." In 2023 3rd International Conference on Consumer Electronics and Computer Engineering (ICCECE), pp. 834-837. IEEE, 2023.DOI: 10.1109/mce.2022.3206678

[8] Wang 2 L. Triboelectric nanogenerators as new energy technology for self- powered systemsand as active mechanical and chemical sensorsJ]. ACS Nano, 2013, 7(11): 9533-9557.

[9] Chang Che, Bo Liu, Shulin Li, Jiaxin Huang, and Hao Hu. Deep learning for precise robot position prediction in logistics. Journal of Theory and Practice of Engineering Science, 3(10):36–41, 2023.DOI: 10.1021/acs.jctc.3c00031.

[10] Hao Hu, Shulin Li, Jiaxin Huang, Bo Liu, and Change Che. Casting product image data for quality inspection with xception and data augmentation. Journal of Theory and Practice of Engineering Science, 3(10):42–46, 2023. https://doi.org/10.53469/jtpes.2023.03(10).06

[11] Nie J, Wang Z, Ren Z, et al. Power generation from the interaction of a liquid droplet and aliquid membrane[J]. Nature Communications, 2019, 10(1): 2264.

[12] Wang S, Lin L, Wang Z L. Nanoscale triboelectric-effect-enabled energy conversion forsustainably powering portable electronics[J]. Nano Letters, 2012, 12(12): 6339-6346.

[13] Chang Che, Qunwei Lin, Xinyu Zhao, Jiaxin Huang, and Liqiang Yu. 2023. Enhancing Multimodal Understanding with CLIP-Based Image-to-Text Transformation. In Proceedings of the 2023 6th International Conference on Big Data Technologies (ICBDT '23). Association for Computing Machinery, New York, NY, USA, 414–418. https://doi.org/10.1145/3627377.3627442

[14] Zhu G, Lin Z H, Jing O, et al. Toward large-scale energy harvesting by a nanoparticle-enhanced triboelectric nanogenerator . Nano Letters, 2013, 13(2): 847-853.

[15] G. Eason, B. Noble, and I. N. Sneddon, "On certain integrals of Lipschitz-Hankel type involving products of Bessel functions," Phil. Trans. Roy. Soc. London, vol. A247, pp. 529–551, April 1955.

[16] PRIETO-AVALOSCRUZRAMOSALORHERNANDEZ G, et al, Wearable devices for physical monitoringof heart: a review [J]. Biosensors, 2022, 12 (5) : 292 - 292.

[17] DRUMMOND C K, LEWANDOWSKI B, MASSARONI CWearable technology for human performance [J]. Frontiersin Physiology , 2022 , 13 : 871159-1 - 871159-2.

[18] CHEN J , ZHU G , YANG W Q,et al Harmonic-resonatorbased triboelectric nanogenerator as a sustainable powersource and a self-powered active vibration sensor [J]. Advanced Materials, 2013, 25 (42) : 6094 - 6099.